\documentclass[aps,prl,twocolumn,superscriptaddress,showpacs,showkeys]{revtex4-2}

\usepackage{color}
\usepackage{graphicx}
\usepackage{dcolumn}
\usepackage{bm}
\usepackage{float}
\usepackage[mathlines]{lineno}
\usepackage{tabularx}
\usepackage[colorlinks,linkcolor=blue,anchorcolor=blue,citecolor=blue,urlcolor=blue,hyperindex,CJKbookmarks]{hyperref}
\usepackage{footnote}
\usepackage{amsmath}
\usepackage{txfonts}
\usepackage{mathptmx}
\usepackage{ragged2e}
\usepackage{booktabs,makecell, multirow, tabularx}
\usepackage{multirow}
\usepackage{braket}
\usepackage{gensymb}
\usepackage{array}
\usepackage{physics} 
\usepackage{textcomp}
\usepackage[T1]{fontenc}

\begin{document}

\title{Strain-driven orbital-selective reconstruction and bicollinear-to-stripe evolution in FeTe}

\author{Zhenfeng Ouyang}\affiliation{Department of Physics, School of Physical Science and Technology, Ningbo University, Ningbo 315211, China}

\author{Yin Chen}\affiliation{School of Physics and Key Laboratory of Quantum State Construction and Manipulation (Ministry of Education), Renmin University of China, Beijing 100872, China}

\author{Yi-Heng Tian}\affiliation{School of Physics and Key Laboratory of Quantum State Construction and Manipulation (Ministry of Education), Renmin University of China, Beijing 100872, China}

\author{Jia-Ming Wang}\affiliation{Center for Materials Theory, Department of Physics and Astronomy, Rutgers University, Piscataway, New Jersey 08854, USA}

\author{Rong-Qiang He}\email{rqhe@ruc.edu.cn}\affiliation{School of Physics and Key Laboratory of Quantum State Construction and Manipulation (Ministry of Education), Renmin University of China, Beijing 100872, China}

\author{Kai Liu}\email{kliu@ruc.edu.cn}\affiliation{School of Physics and Key Laboratory of Quantum State Construction and Manipulation (Ministry of Education), Renmin University of China, Beijing 100872, China}

\author{Zhong-Yi Lu}\email{zlu@ruc.edu.cn}\affiliation{School of Physics and Key Laboratory of Quantum State Construction and Manipulation (Ministry of Education), Renmin University of China, Beijing 100872, China}\affiliation{Hefei National Laboratory, Hefei 230088, China}

\date{\today}

\begin{abstract}
FeTe, as a representative parent material among iron-based superconductors, provides an ideal platform for exploring the interplay among orbital-selective correlations, magnetism, and unconventional superconductivity. However, a unified picture of the correlated electronic structure and magnetism of FeTe under strain remains to be fully clarified. Here, combining density functional theory plus dynamical mean-field theory and Heisenberg model analysis, we uncover an orbital-selective reconstruction of the correlated electronic structure and reveal a strain-driven trajectory from bicollinear to stripe antiferromagnetism (AFM) via an intermediate competing staggered $n$-mer AFM regime in FeTe. Moderate strain gives rise to a regime where more coherent quasiparticles coexist with suppressed local moments. Further strain drives FeTe into an incoherent correlated regime with robust local moments and Fe-$3d_{z^2}$-dominated low-energy states. These results establish a strain-driven trajectory across distinct magnetic and correlated electronic states in FeTe.
\end{abstract}


\maketitle

\textit{Introduction—}Since the discovery of iron-based superconductors, these materials have emerged as a prominent platform for exploring the intimate interplay among magnetism, electronic correlations, orbital degrees of freedom, and superconductivity. FeTe is one of the most representative parent materials among iron-based superconductors~\cite{PhysRevLett.102.177003,PhysRevLett.102.247001,PhysRevB.84.064403,PhysRevB.90.165123,Maheshwari2015,PhysRevB.100.054405,He2014,Liang2020,Yi2023,Yi2024,Yan2026AM,PhysRevLett.104.017003,PhysRevB.78.224503,Mizuguchi2009}. The magnetic ground state of bulk FeTe is a bicollinear antiferromagnetic (AFM) order~\cite{PhysRevLett.102.177003,PhysRevLett.102.247001,PhysRevB.84.064403,PhysRevB.90.165123,Maheshwari2015,PhysRevB.100.054405} rather than the stripe AFM phase common to many other iron pnictides~\cite{delaCruz2008,PhysRevLett.101.257003,PhysRevB.78.064515}. This unusual magnetic order has long been attributed to the delicate competition between magnetic exchange interactions and strong Hund's correlations~\cite{PhysRevLett.102.177003,Haule2009,Yin2011}, making FeTe an ideal platform for exploring the interplay between orbital-selective electronic correlation and magnetic instability.

Recent experimental reports~\cite{Yan2026,Xu2026} have advanced our understanding of FeTe: the stoichiometric FeTe thin films grown on SrTiO$_3$ substrates have been found to become superconducting once excess interstitial Fe is removed, suggesting a close connection between the suppression of bicollinear AFM order and the appearance of superconductivity under in-plane tensile strain. Meanwhile, another experimental report~\cite{xffv-kpjp} reveals strain-induced orbital reconstruction, characterized by a transfer of spectral weight between the Fe-3$d_{x^2-y^2}$ and Fe-3$d_{z^2}$ orbitals under tensile strain. These findings suggest that tensile strain not only suppresses magnetism but also substantially reconstructs the correlated electronic structure of FeTe. 

Although existing studies~\cite{Yan2026,Xu2026,Ciechan2013} show that moderate tensile strain can suppress bicollinear AFM order of FeTe and induce superconductivity, the complete strain-driven trajectory connecting the competing bicollinear AFM, staggered $n$-mer AFM~\cite{PhysRevB.93.205154,PhysRevB.91.020504}, and stripe AFM, together with the accompanying orbital-selective reconstruction, remains incomplete. Addressing this question is crucial for establishing the complete magnetic phase diagram of strained FeTe and for understanding how orbital reconstruction reshapes magnetic interactions in this Hund's metal.

In this Letter, we employ density functional theory plus dynamical mean-field theory (DFT+DMFT) calculations combined with Heisenberg model studies to investigate the evolution of correlated electronic and magnetic properties of FeTe under in-plane biaxial tensile strain. We demonstrate that tensile strain drives a pronounced orbital-selective reconstruction of the correlated electronic structure. Moreover, we reveal that the bicollinear AFM order is not simply suppressed by strain; biaxial tension instead drives a trajectory toward a stripe AFM ground state through an intermediate competing staggered $n$-mer AFM regime characterized by coherence-enhanced quasiparticles and suppressed local moments. Our results establish a unified microscopic picture connecting the correlated electronic structure and magnetism of strained FeTe.

\begin{figure*}[t]
\centering
\includegraphics[width=17.2cm]{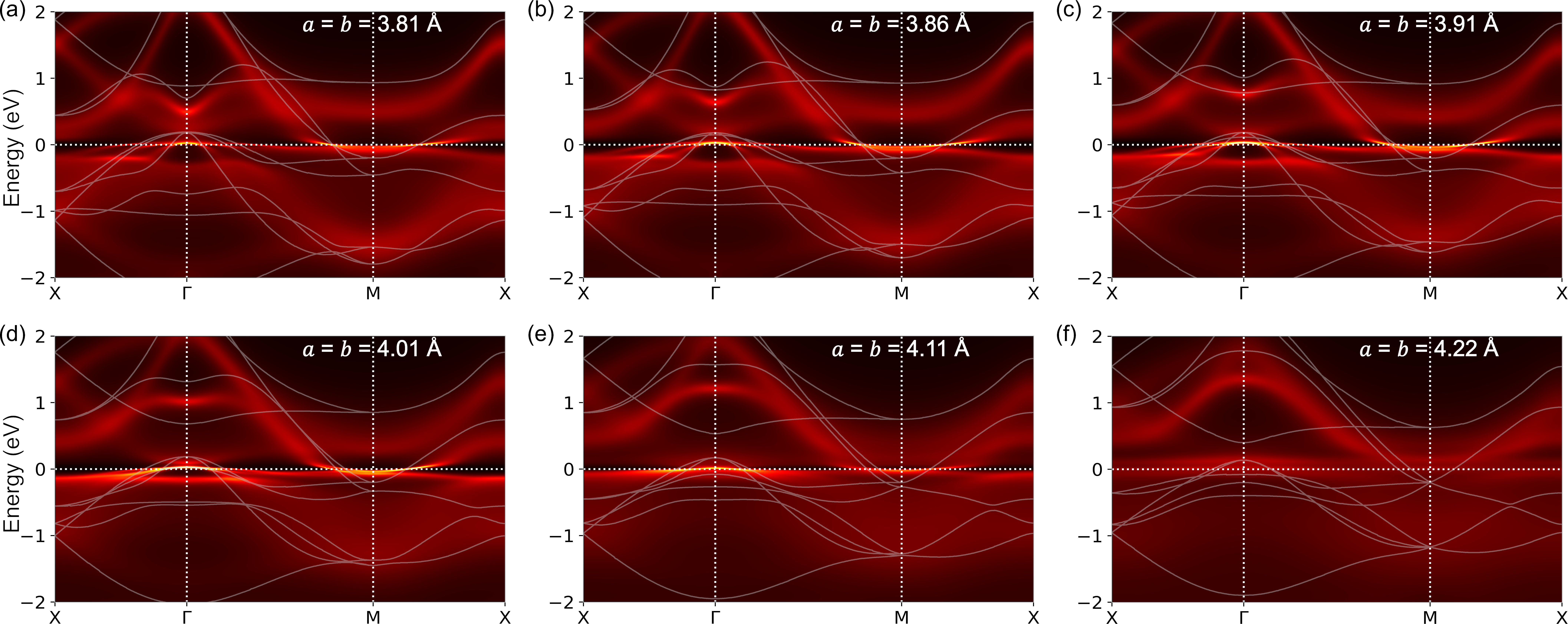}
\caption{DFT+DMFT calculated spectral functions $A(\bf{k}, \omega)$ (temperature $T$ = 100 K) and the DFT calculated band structures (grey lines) for different in-plane biaxial tensile-strained FeTe. The Fermi level is set as zero.}
\label{fig:Fig1}
\end{figure*}

\textit{Strain-driven orbital-selective reconstruction of electronic structure—}To elucidate the evolution of the correlated electronic structure under tensile strain, we perform DFT+DMFT calculations on paramagnetic tetragonal FeTe (space group $P4/nmm$) with systematically varied in-plane lattice constants. As shown in Fig.~\ref{fig:Fig1}, the momentum-resolved spectral functions $A(\bf{k}, \omega)$ are presented together with the corresponding DFT band structures for comparison. The DFT results suggest that FeTe is a multi-orbital metal, with three hole pockets around the $\Gamma$ point and two electron pockets around the $M$ point. Upon incorporating electronic correlations within the DFT+DMFT framework, the quasiparticle bandwidth is significantly renormalized, accompanied by pronounced spectral broadening and suppressed quasiparticle coherence. These characteristics are consistent with the established consensus that FeTe is a strongly correlated Hund's metal~\cite{Haule2009,Yin2011,PhysRevB.88.115130}. Having established the overall correlated electronic structure of FeTe, we next investigate its evolution under increasing biaxial tensile strain.

For the first four in-plane lattice constants ($a$ = 3.81, 3.86, 3.91, and 4.01 {\AA}), we find that the increasing in-plane tensile strain enhances the spectral weight and coherence in the vicinity of the Fermi level. However, upon further increasing the tensile strain ($a$ = 4.11 and 4.22 {\AA}), a different evolution emerges, in which the low-energy spectral weight and the quasiparticle coherence are gradually suppressed, as shown in Figs.~\ref{fig:Fig1}(e)-(f). Moreover, the quasiparticle spectrum of the system also undergoes an orbital-selective reconstruction. In particular, the bottom of the conduction band at the $\Gamma$ point that is dominated by the Fe-3$d_{x^2-y^2}$ orbital exhibits strong spectral weight for the small-strain cases [Figs.~\ref{fig:Fig1}(a)-(d)], whereas its spectral weight is progressively suppressed as the tensile strain increases, and eventually becomes barely discernible in the large-strain regime [Figs.~\ref{fig:Fig1}(e) and (f)]. 

\begin{figure}[b]
\centering
\includegraphics[width=8.6cm]{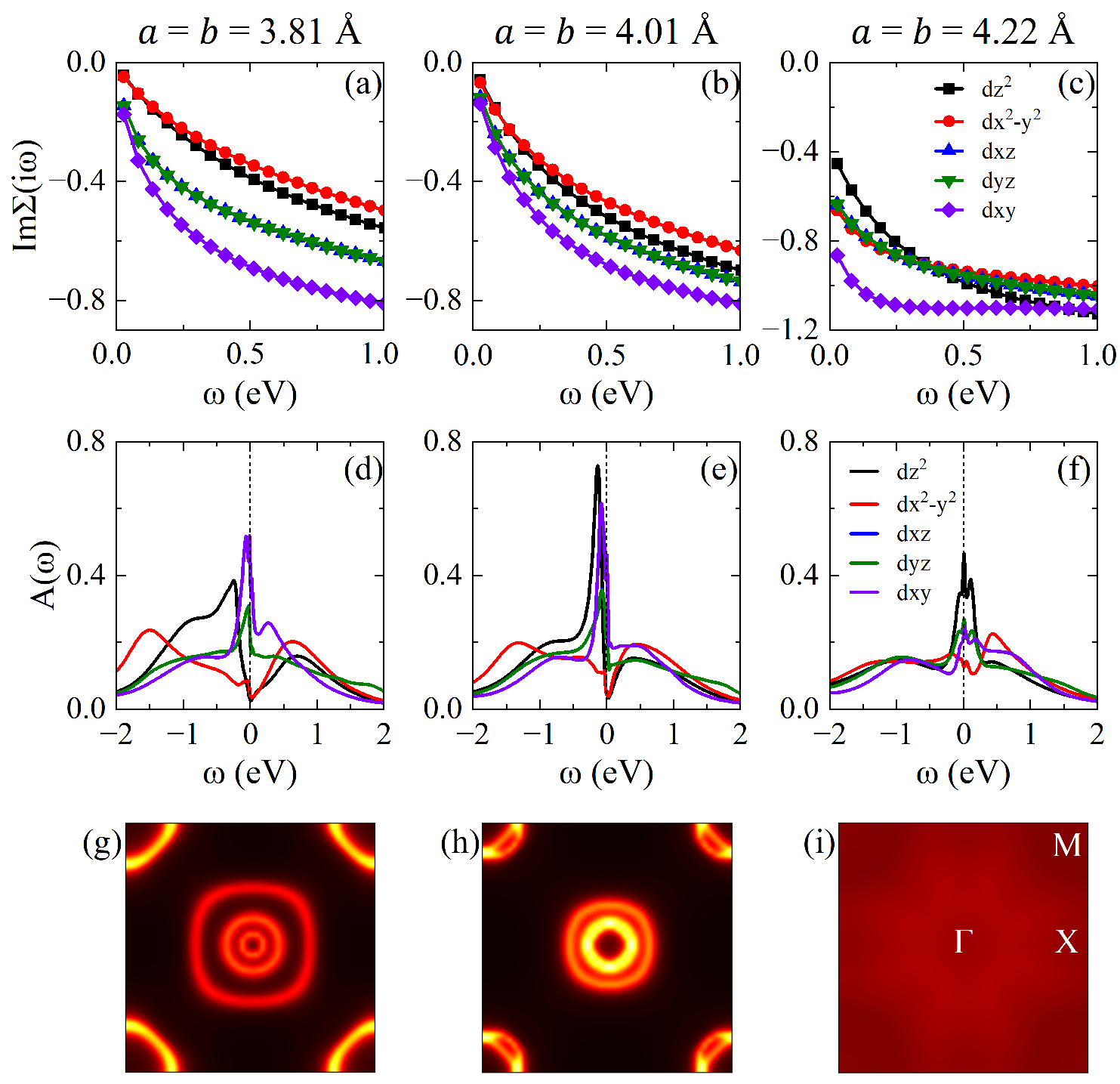}
\caption{DFT+DMFT calculated (temperature $T$ = 100 K) (a)-(c) imaginary parts of Matsubara self-energy functions Im$\Sigma(i\omega)$, (d)-(f) spectral functions $A(\omega)$, and (g)-(i) Fermi surfaces at the $k_z$ = 0 plane for in-plane tensile-strained FeTe with crystal lattice constants $a$ = 3.81, 4.01, and 4.22 {\AA}, respectively.}
\label{fig:Fig2}
\end{figure}

The above results reveal a two-stage evolution of the correlated electronic states: moderate tensile strain enhances quasiparticle coherence, marking a crossover from the incoherent regime of FeTe toward a more Fermi-liquid-like state. In contrast, larger tensile strain drives the orbital-selective reconstruction of the correlated electronic structure. To uncover its microscopic origin, we further analyze the self-energy functions Im$\Sigma(i\omega)$, spectral functions $A(\omega)$, and Fermi surfaces obtained from DFT+DMFT calculations. Here, as shown in Fig.~\ref{fig:Fig2}, we focus on three representative cases ($a$ = 3.81, 4.01, and 4.22 Å), which serve as three key stages along the nonmonotonic evolution of the correlated electronic structure of tensile-strained FeTe.

We first study the orbital-dependent electronic correlation through the imaginary part of the Matsubara self-energy Im$\Sigma(i\omega)$, as shown in Figs.~\ref{fig:Fig2}(a)-(c). For the case of $a$ = 3.81 {\AA}, the Fe-3$d$ orbitals exhibit orbital-selective electronic correlation, which is a typical signature of Hund's metal~\cite{PhysRevLett.102.177003,Haule2009,Yin2011,PhysRevB.109.115114,PhysRevB.109.165140,PhysRevB.102.125130}. The self-energy functions of the Fe-$t_{2g}$ orbitals show a sharper asymptotic behavior at the low-frequency region than those of the Fe-$e_g$ orbitals, suggesting that the correlation of the Fe-$t_{2g}$ orbitals is stronger. Upon applying moderate tensile strain ($a$ = 4.01 {\AA}), we find that the low-frequency intercept of Im$\Sigma(i\omega)$ for the Fe-$t_{2g}$ orbitals moves closer to zero, indicating a reduced incoherent scattering rate and an enhanced tendency toward Fermi-liquid behavior. This result is consistent with the evolution that is exhibited in Figs.~\ref{fig:Fig1}(a)-(d), where moderate tensile strain can enhance the spectral weight and coherence of the spectra. In sharp contrast, when the tensile strain is further increased to $a$ = 4.22 {\AA}, the low-frequency intercept of Im$\Sigma(i\omega)$ increases markedly for all five Fe-3$d$ orbitals, indicating a substantial enhancement of incoherent electronic scattering. This is a non-Fermi liquid behavior. In particular, the Fe-3$d_{xy}$ orbital not only retains the largest low-frequency intercept of Im$\Sigma(i\omega)$, but also becomes increasingly separated from the other orbitals, suggesting a further enhancement of orbital-selective electronic correlations in the large-strain regime.

The orbital-resolved spectral functions shown in Figs.~\ref{fig:Fig2}(d)-(f) further reveal the microscopic origin of the strain-induced electronic reconstruction. From $a$ = 3.81 to 4.01 {\AA}, the low-energy spectral weight of both the Fe-3$d_{xy}$ and Fe-3$d_{z^2}$ orbitals increases significantly. At the same time, an apparent redistribution of low-energy spectral weight from the Fe-3$d_{xy}$ to the Fe-3$d_{z^2}$ orbital is observed, as evidenced by the enhanced Fe-3$d_{z^2}$-derived quasiparticle peak and its shift toward the Fermi level. Upon further increasing the tensile strain to $a$ = 4.22 {\AA}, however, the quasiparticle spectral weight of all five Fe-3$d$ orbitals is suppressed, reflecting the reduced coherence of the correlated electronic states. Meanwhile, the Fe-3$d_{z^2}$ orbital provides the dominant contribution to the electronic states near the Fermi level, suggesting a fundamental reconstruction of the low-energy correlated electronic structure. 

The corresponding evolution is also directly reflected in the Fermi surface topology [Figs.~\ref{fig:Fig2}(g)-(i)]. Well-defined hole pockets around the $\Gamma$ point and electron pockets around the $M$ point are preserved under small and moderate tensile strain, while the large-strain regime exhibits a dramatic loss of coherent Fermi-surface spectral weight, as well as a complete reconstruction of band topology. These results strongly indicate that the nonmonotonic evolution of the electronic structure identified in Fig.~\ref{fig:Fig1} originates from a strain-induced orbital-selective reconstruction of the correlated electronic states, in which the Fe-3$d_{xy}$ orbital plays a crucial role in the coherence of the correlated electronic structure.

The orbital reconstruction originates from strain-modified Fe-Te hybridization, which selectively affects the Fe-3$d_{xy}$ states strongly coupled with Te-5$p$ orbitals. Under tensile strain, the local Fe-Te coordination environment evolves substantially: the Fe-Te bond length increases from 2.49 {\AA} at $a$ = 3.81 {\AA} to 2.50 {\AA} and 2.52 {\AA} at $a$ = 4.01 {\AA} and 4.22 {\AA}, respectively, while the in-plane Te-Fe-Te bond angle expands from 100.09$^\circ$ to 106.82$^\circ$ and 113.72$^\circ$. Such structural evolution modifies the orbital-dependent hybridization between Fe-3$d$ and Te-5$p$ states, leading to a redistribution of low-energy spectral weight among Fe orbitals. In particular, the relative weakening of Fe-3$d_{xy}$ contribution and enhancement of Fe-3$d_{z^2}$ contribution eventually drive the orbital-selective reconstruction toward a Fe-3$d_{z^2}$-dominated low-energy electronic structure under large tensile strain.

\textit{Strain-driven magnetic transition—}The observed emergence of incoherent Fermi surfaces in large tensile-strained FeTe suggests that the correlated electronic state of the system gradually evolves toward a more localized regime. Especially in Hund's metals, the suppression of quasiparticle coherence is generally accompanied by the formation of robust local magnetic moments~\cite{Yin2011,PhysRevB.109.115114,Deng2019,Stadler2019}. Such a coherence-incoherence crossover of electronic structure is expected to modulate the local moments and the competition among magnetic orders in FeTe.

\begin{figure}[t]
\centering
\includegraphics[width=8.6cm]{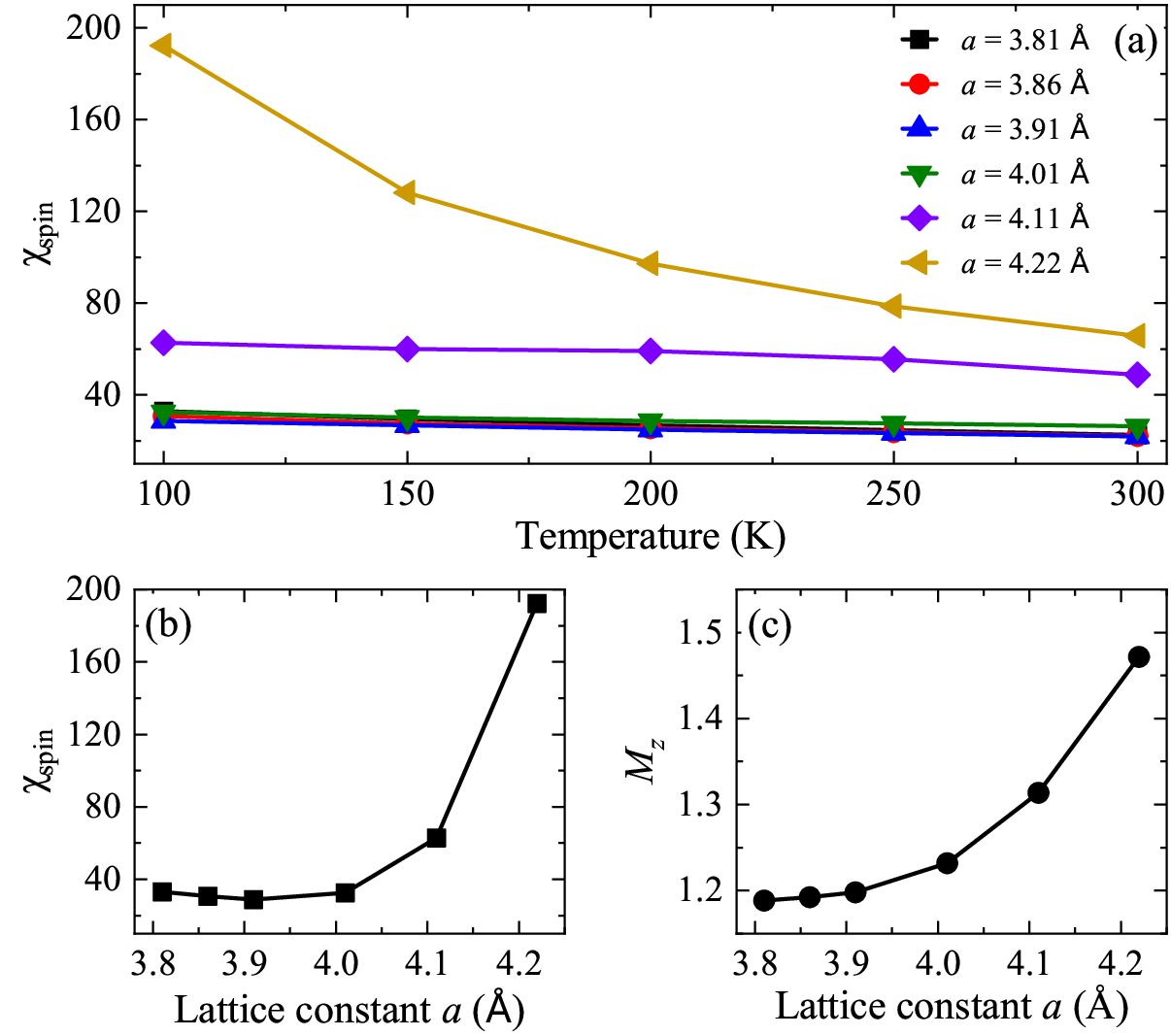}
\caption{(a) Evolution of static local spin susceptibilities $\chi_{\rm{spin}}$ of FeTe with temperature under different strain. (b) Evolution of static local spin susceptibilities $\chi_{\rm{spin}}$ of FeTe at 100 K with in-plane tensile strain. (c) Evolution of effective local spin moments $M_z$ of FeTe with in-plane tensile strain.}
\label{fig:Fig3}
\end{figure}

To clarify how tensile strain affects the magnetism of FeTe, we first study the static local spin susceptibilities $\chi_{\rm{spin}}$ of tensile-strained paramagnetic FeTe under different temperatures, as shown in Fig.~\ref{fig:Fig3}(a). For $a$ =3.81 to 4.01 {\AA}, the $\chi_{\rm{spin}}$ exhibits a weak temperature dependence. To better reveal its evolution, we further compare the $\chi_{\rm{spin}}$ at 100 K, as exhibited in Fig.~\ref{fig:Fig3}(b). We find that the susceptibility also displays a nonmonotonic dependence on tensile strain, showing a slight decrease from $a$ = 3.81 to 3.91 {\AA}, followed by a gradual increase under larger strains. Combined with the evolution of the correlated electronic structure shown in Figs.~\ref{fig:Fig1} and~\ref{fig:Fig2}, a unified physical picture of tensile-strained FeTe is revealed: moderate tensile strain suppresses local-moment formation and stabilizes a more coherent correlated electronic structure, whereas further increasing the tensile strain drives the system toward a local-moment-dominated regime characterized by a Curie-Weiss-like temperature dependence of the local susceptibility.

\begin{table}[t]
\centering
\caption{Relative energies (meV/f.u.) of competing magnetic configurations in tensile-strained FeTe for different in-plane lattice constants. The energies of the lowest-energy magnetic state for each lattice constant are set to zero. 
For $a$ = 4.11 and 4.22 {\AA}, the ferromagnetic (FM) configuration is unstable and relaxes to a nonmagnetic (NM) solution during self-consistent calculations.}
\label{tab:tab1}
\begin{tabular*}{\columnwidth}{@{\extracolsep{\fill}}cccccccc}
\toprule
$a$ (\AA) & NM & FM & Néel & stripe & bicollinear & Dimer & Trimer \\
\midrule
3.81 & 186.33 & 84.52 & 71.49 & 21.17 & \bfseries 0.00 & 4.37 & 9.10 \\
3.86 & 187.26 & 114.39 & 64.78 & 17.37 & 6.16 & \bfseries 0.00 & 4.15 \\
3.91 & 192.78 & 189.40 & 63.26 & 16.06 & 14.63 & \bfseries 0.00 & 3.35 \\
4.01 & 210.50 & 210.22 & 62.04 & 7.39 & 32.43 & \bfseries 0.00 & 0.54 \\
4.11 & 243.96 & - & 68.07 & \bfseries 0.00 & 51.12 & 4.84 & 2.14 \\
4.22 & 301.01 & - & 84.65 & \bfseries 0.00 & 77.32 & 19.86 & 13.74 \\
\bottomrule
\end{tabular*}
\end{table}

To further verify the physical picture proposed above, we introduce an effective local spin moment $M_z$ = $\sqrt{\langle{S_z^2}\rangle}$ to quantitatively characterize the evolution of local spin configurations under tensile strain, which is calculated from the local spin multiplets (Table~\ref{tab:tab2}). Unlike the magnetic susceptibility, which measures the magnetic response, $M_z$ directly reflects the stability of local spin moments: a large $M_z$ indicates that the statistical weight is increasingly concentrated on high-spin local multiplets, corresponding to a more stabilized local-moment configuration. Similar analysis has been successfully utilized in the study on Ruddlesden-Popper nickelates~\cite{1412-nfzm}. As shown in Fig.~\ref{fig:Fig3}(c), $M_z$ exhibits a gradual increase under small and moderate tensile strains, followed by a much sharper enhancement in the large-strain regime. This result indicates the formation of robust local magnetic moments. 

Such a pronounced enhancement of local moments naturally raises the question of whether the magnetic ground state itself is reconstructed under large tensile strain. In Table~\ref{tab:tab1}, we summarize the DFT-calculated energies of different magnetic configurations of FeTe for all considered in-plane lattice constants. For the equilibrium lattice constant ($a=3.81$ {\AA}), corresponding to the theoretically optimized bulk FeTe, the bicollinear AFM order is found to be the ground state, in agreement with numerous previous studies~\cite{PhysRevLett.102.177003,PhysRevLett.102.247001,PhysRevB.84.064403,PhysRevB.90.165123,Maheshwari2015,PhysRevB.100.054405}. Upon increasing the tensile strain to the intermediate regime ($a$ = 3.86 to 4.01 {\AA}), the energy difference between the Dimer and Trimer AFM orders is progressively reduced, although the Dimer AFM order remains energetically favorable. The enhanced near-degeneracy between the Dimer and Trimer AFM orders is indicative of an increasingly competing magnetic landscape. Intriguingly, this specific intermediate strain range, where magnetic competition is enhanced and quasiparticle coherence is significantly promoted (Figs.~\ref{fig:Fig1} and~\ref{fig:Fig2}), coincides with the regime in which superconductivity has been recently reported in tensile-strained stoichiometric FeTe thin films~\cite{Yan2026,Xu2026}. In the large-strain regime ($a$ = 4.22 {\AA}), stripe AFM order becomes significantly lower in energy than other magnetic configurations, demonstrating that tensile strain ultimately stabilizes a stripe AFM ground state in FeTe. This result is consistent with the DFT+DMFT results shown in Figs.~\ref{fig:Fig1}(f) and ~\ref{fig:Fig2}(i), where the quasiparticle spectra exhibit incoherence due to the robust local moments.

\begin{figure}[t]
\centering
\includegraphics[width=8cm]{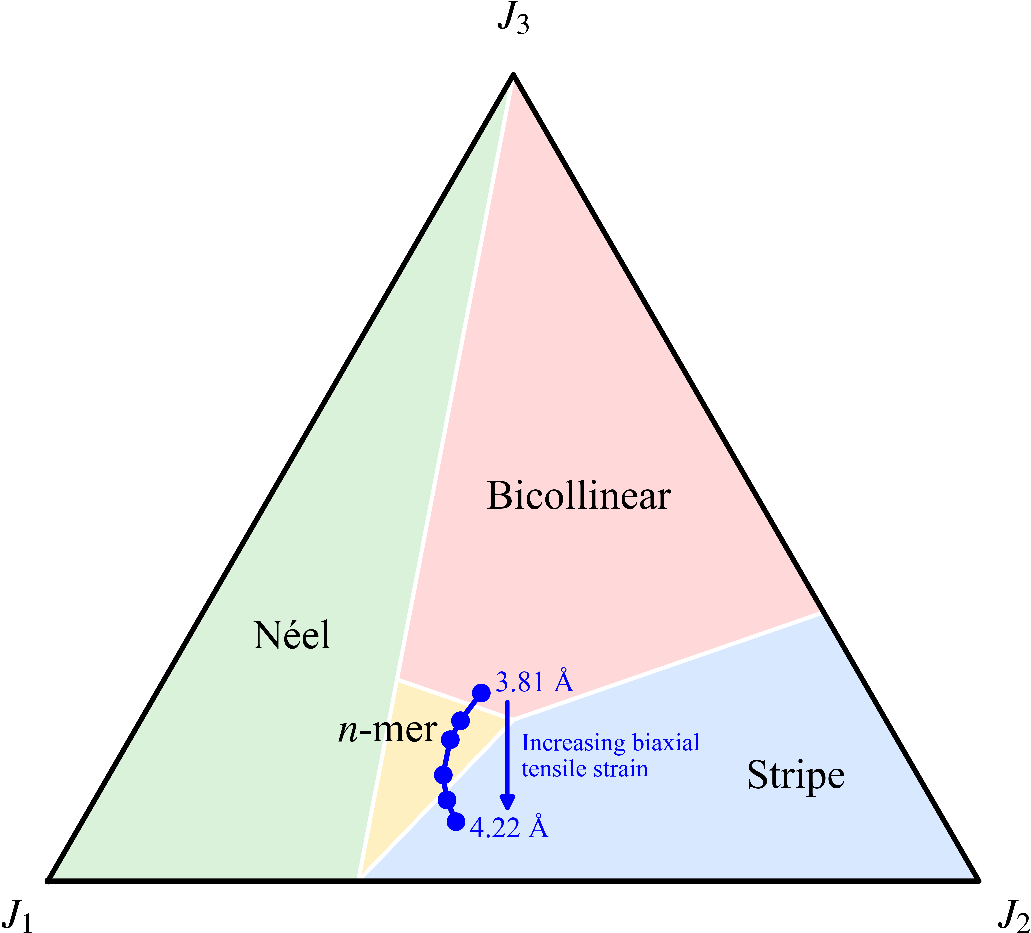}
\caption{Phase diagram of the $J_1$-$J_2$-$J_3$ Heisenberg model for FeTe. The phase boundaries separating the Néel, stripe, bicollinear, and staggered $n$-mer AFM phases are determined from the results of Ref.~\cite{PhysRevLett.102.177003,PhysRevB.93.205154,PhysRevB.91.020504}. The blue symbols denote the normalized exchange couplings extracted from DFT calculations for FeTe under different in-plane biaxial tensile strains.}
\label{fig:Fig4}
\end{figure}

We further map the DFT total energies onto an effective $J_1$-$J_2$-$J_3$ Heisenberg model and project the extracted exchange interactions onto the established magnetic phase diagram~\cite{PhysRevLett.102.177003,PhysRevB.93.205154,PhysRevB.91.020504}, as shown in Fig.~\ref{fig:Fig4}. The bicollinear AFM phase is energetically stabilized when $J_3 > J_2/2$ and $J_2 > J_1/2$, whereas the stripe AFM phase is favored once $J_2 > J_1/2$ but $J_3 < J_2/2$~\cite{PhysRevLett.102.177003}. When $J_1$ - 2$J_2$ + 2$J_3$ > 0, the staggered $n$-mer AFM state will be energetically lower than the stripe AFM state~\cite{PhysRevB.93.205154,PhysRevB.91.020504}. Remarkably, the extracted exchange interactions under different tensile strains (blue symbols in Fig.~\ref{fig:Fig4}) not only conform to the established $J_1$-$J_2$-$J_3$ phase diagram, but also reveal a well-defined strain-driven evolution trajectory on the phase diagram. At the equilibrium lattice constant ($a$ = 3.81 {\AA}), FeTe lies inside the bicollinear AFM region. The intermediate-strain cases ($a$ = 3.86 to 4.01 {\AA}) are all located in the staggered $n$-mer AFM states, which lie in the vicinity of the bicollinear-stripe phase boundary and strong magnetic competition is expected as shown in Table~\ref{tab:tab1}. Upon further increasing the tensile strain, the trajectory enters the stripe AFM region, in agreement with our DFT results showing that the stripe AFM state becomes the lowest-energy magnetic configuration.

Finally, we establish a unified microscopic picture of the electronic and magnetic properties that is closely connected through the Heisenberg phase diagram and the DFT+DMFT results. FeTe, as a typical Hund's metal, hosts bicollinear AFM order with robust local spin moments stabilized by Hund's coupling. Moderate tensile strain drives the system into the regime of staggered $n$-mer AFM, in which the energy of Dimer and Trimer AFM states remain nearly degenerate, indicating strong magnetic competition. Meanwhile, the local spin moments are destabilized and the coherence of the low-energy quasiparticle spectra is significantly improved. Upon further increasing the tensile strain, robust local spin moments are recovered as the system crosses over to the stripe AFM phase, accompanied by a complete orbital-selective reconstruction of the correlated electronic structure, in which the Fe-$3d_{z^2}$ orbital dominates the low-energy electronic states.

\textit{Discussion and Conclusion—}Although previous studies have established the magnetism of bulk FeTe, how lattice strain modifies the magnetism and the associated correlated electronic states of FeTe remains incomplete. Our work establishes a comprehensive physical picture of the strain-driven evolution in FeTe, revealing the intimate interplay between orbital-selective electronic structure reconstruction and magnetic phase evolution. In particular, our work reveals an intermediate regime characterized by enhanced magnetic competition and coherent quasiparticles. Remarkably, this regime overlaps with the strain window where superconductivity has been reported in recent experiments on stoichiometric FeTe thin films~\cite{Yan2026,Xu2026}. In addition, our work shows the potential of strain engineering as a route for manipulating correlated quantum phases in FeTe, and we further suggest that other similar types of crystal modulations~\cite{xu2026selectivestabilizationantiferromagneticorders}, such as uniaxial strain or pressure, may also provide opportunities to explore and manipulate the emergent correlated phases in FeTe, which await future experimental and theoretical investigations.

More broadly, our findings highlight a conceptual distinction between the unconventional superconductivity in strained FeTe and hole-doped cuprates. In hole-doped cuprates, superconductivity generally emerges between an AFM Mott insulating phase and an overdoped Fermi-liquid regime. In contrast, the intermediate regime of tensile-strained FeTe is not characterized by the destruction of a Mott gap, but rather by the competing AFM phases with coherent electronic states. This difference suggests that unconventional superconductivity may arise from diverse electronic environments and competing energy scales, while strong magnetic fluctuations stand out as a common thread linking these seemingly distinct unconventional superconducting systems.

In summary, our work uncovers that FeTe is a promising platform to study the interplay between Hund's electronic correlation, non-Fermi liquid behavior, and magnetism, in which strain serves as an effective tuning parameter to drive the system across distinct correlated electronic and magnetic states. In-plane biaxial tensile strain drives the magnetic ground state from the bicollinear AFM phase to the stripe AFM phase through an intermediate regime of competing staggered $n$-mer orders. The combined DFT+DMFT calculations and Heisenberg-model analysis demonstrate that this magnetic evolution is intimately coupled with an orbital-selective reconstruction of the correlated electronic structure. Moderate tensile strain suppresses the formation of local spin moments, creating a regime where the coherent electronic structure coexists with strong magnetic competition. In the large-strain regime, FeTe undergoes a simultaneous magnetic and electronic reconstruction: the magnetic ground state is stabilized as the stripe AFM order, and the correlated electronic structure develops an orbital-selective reconstruction characterized by Fe-$3d_{z^2}$-dominated low-energy electronic states.

\textit{Note added—}During the preparation of this manuscript, we noticed another work~\cite{hua2026competingextendedsdwavepairing} that studied the superconducting pairing of electron-doped stoichiometric FeTe.

\textit{Acknowledgments—}This work was supported by the National Natural Science Foundation of China (Grant No. 12434009) and the National Key R\&D Program of China (Grants No. 2024YFA1408601 and No. 2024YFA1408602). 
K.L. was supported by the National Key R\&D Program of China (Grant No. 2022YFA1403103).
Z.Y.L. was also supported by the Innovation Program for Quantum Science and Technology (Grant No. 2021ZD0302402). 
Computational resources were provided by the Physical Laboratory of High Performance Computing in Renmin University of China.

\bibliography {FeTe}

\appendix

\section{End Matter}

\textit{Method of calculations—}The crystal structures of different strained bulk FeTe were optimized by using the QUANTUM-ESPRESSO package~\cite{Giannozzi2009}, in which only the in-plane lattice constants were fixed. The magnetic properties of different strained bulk FeTe were calculated by using the VASP package~\cite{PhysRevB.47.558,Kresse1996,PhysRevB.54.11169}. The nonmagnetic, ferromagnetic, Néel AFM, stripe AFM, bicollinear AFM, and Dimer AFM states were studied with a 2$\sqrt{2}$ $\times$ 2$\sqrt{2}$ supercell. The Trimer AFM state was studied with a 3$\sqrt{2}$ $\times$ 3$\sqrt{2}$ supercell. The generalized gradient approximation of Perdew-Burke-Ernzerhof type for the exchange-correlation potentials~\cite{PhysRevLett.77.3865} was adopted in all DFT calculations. The DFT parts of our DFT+DMFT calculations are performed by the WIEN2K code~\cite{Blaha2020} with the full-potential linearized augmented plane-wave method. The EDMFTF software package~\cite{PhysRevB.81.195107} is used to perform the charge fully self-consistent DFT+DMFT calculations. The systems are enforced to be paramagnetic. The Fe-3$d$ orbitals were treated as correlated orbitals, and a single impurity problem was constructed because the two Fe atoms are equivalent within the paramagnetic unit cell. The local coordinate system of Fe atoms was defined with the $x$ and $y$ axes aligned along the Fe-Fe bond directions, while the $z$ axis was kept parallel to the global crystallographic $z$ axis. The Fe-$3d$ orbitals were further defined based on this local coordinate frame. The Coulomb interaction parameter $U$ was set to be 5.0 eV, and the Hund’s exchange parameter $J_H$ was set to be 0.88 eV, which are typical values for 3$d$-transition elements. The density-density form of the Coulomb repulsion is used. The finite-temperature quantum impurity problems within the DMFT framework are solved by the hybridization expansion continuous-time quantum Monte Carlo (CTQMC) impurity solver~\cite{PhysRevB.75.155113} with an exact double-counting scheme~\cite{PhysRevLett.115.196403} for the self-energy function. The real-frequency self-energy function was obtained by using the maximum entropy method analytical continuation~\cite{Jarrell1996}, then was used to calculate the momentum-resolved spectral function and the other related physical quantities.

In our Heisenberg-model analysis, we considered the nearest, the next-nearest, and the next-next-nearest neighbor couplings $J_1$, $J_2$, and $J_3$, 
\begin{equation}
H = 
J_1 \sum_{\langle ij\rangle} \mathbf{S}_i \cdot \mathbf{S}_j
+
J_2 \sum_{\langle\langle ij\rangle\rangle} \mathbf{S}_i \cdot \mathbf{S}_j
+
J_3 \sum_{\langle\langle\langle ij\rangle\rangle\rangle} 
\mathbf{S}_i \cdot \mathbf{S}_j.
\end{equation}
$\langle ij\rangle$, $\langle\langle ij\rangle\rangle$, and $\langle\langle\langle ij\rangle\rangle\rangle$ denote the summation over the nearest, next-nearest, and next-next-nearest neighbors, respectively. By assuming that the energy differences between strained FeTe with different magnetic orderings arise predominantly from the interactions between the Fe moments with spin $\mathbf{S}$, the effective exchange parameters $J_1$, $J_2$, and $J_3$ can be solved. The effective exchange parameters $J_1$, $J_2$, and $J_3$ are calculated based on the energy of Néel, stripe, bicollinear, and Dimer AFM states.

\begin{table}[h]
\centering
\caption{DFT+DMFT calculated weights (\%) of the Fe-3$d$ orbitals local spin multiplets for different in-plane tensile strained FeTe at 100 K. The good quantum numbers $N$ and $S_z$ denote the total occupancy and total spin of the Fe-3$d$ orbitals, respectively, which are used to label different local spin states.}
\label{tab:tab2}
\begin{tabular*}{\columnwidth}{@{\extracolsep{\fill}}c ccc ccc cc c}
\toprule
$N$ 
& \multicolumn{3}{c}{5} 
& \multicolumn{3}{c}{6} 
& \multicolumn{2}{c}{7}
& other \\
\cmidrule(lr){2-4}
\cmidrule(lr){5-7}
\cmidrule(lr){8-9}
$S_z$
& 0.5 & 1.5 & 2.5
& 0 & 1 & 2
& 0.5 & 1.5
& other \\
\midrule
3.81 {\AA}
& 6.10 & 5.22 & 1.37
& 9.31 & 19.78 & 12.23
& 19.08 & 17.25
& 9.68 \\
\midrule
3.86 {\AA}
& 5.91 & 5.11 & 1.36
& 9.21 & 19.68 & 12.42
& 19.12 & 17.58
& 9.62 \\
\midrule
3.91 {\AA}
& 5.81 & 5.08 & 1.37
& 9.13 & 19.64 & 12.66
& 19.02 & 17.78
& 9.52 \\
\midrule
4.01 {\AA}
& 5.31 & 5.08 & 1.49
& 7.69 & 19.37 & 14.02
& 18.14 & 19.01
& 9.91 \\
\midrule
4.11 {\AA}
& 4.35 & 5.30  & 1.87
& 4.19 & 18.60 & 17.43
& 15.40 & 21.73
& 11.14 \\
\midrule
4.22 {\AA}
& 2.78 & 5.86 & 2.78
& 3.55 & 16.53 & 24.95
& 9.98 & 26.62
& 6.97 \\
\bottomrule
\end{tabular*}
\end{table}

\textit{Local spin multiplets—}As shown in Table~\ref{tab:tab2}, we list the local spin multiplets of Fe-3$d$ orbitals and their corresponding weights for different in-plane biaxial tensile strained FeTe systems, which are obtained from the CTQMC impurity solver and reflect valence fluctuations. We calculate the local spin moments $M_z$ = $\sqrt{\langle{S_z^2}\rangle}$ [See Fig.~\ref{fig:Fig3}(c)] to further quantify their stability. Similar analysis has been successfully applied in Ruddlesden-Popper nickelates~\cite{PhysRevB.109.115114,1412-nfzm,PhysRevB.109.165140,PhysRevB.111.125111}, where, together with experimental observations, it has been shown to be useful for identifying the evolution among different electronic phases, including spin-density-wave/AFM, superconducting, and Fermi-liquid regimes.

\end{document}